# Simple self-gettering differential-pump for minimizing source oxidation in oxide-MBE environment

Yong-Seung Kim[1], Namrata Bansal[2], and Seongshik Oh[1, a)]

[1] Department of Physics & Astronomy, Rutgers, The State University of New Jersey, 136 Frelinghuysen Rd, Piscataway, NJ, 08854, U.S.A.

[2] Department of Electrical and Computer Engineering, Rutgers, The State University of New Jersey, 94 Brett Rd, Piscataway, NJ, 08854, USA

a) Electronic mail: ohsean@physics.rutgers.edu

MATERIAL NAMES: Strontium (Sr), Strontium Oxide (SrO), Oxygen ($O_2$)

# Abstract

Source oxidation of easily oxidizing elements such as Ca, Sr, Ba, and Ti in an oxidizing ambient leads to their flux instability and is one of the biggest problems in the multi-elemental oxide Molecular Beam Epitaxy technique. Here we report a new scheme that can completely eliminate the source oxidation problem: a self-gettering differential pump using the source itself as the pumping medium. The pump simply comprises a long collimator mounted in front of the source in extended port geometry. With this arrangement, the oxygen partial pressure near the source was easily maintained well below the source oxidation regime, resulting in a stabilized flux, comparable to that of an ultra-high-vacuum environment. Moreover, this pump has a self-feedback mechanism that allows a stronger pumping effectiveness for more easily oxidizing elements, which is a desired property for eliminating the source oxidation problem.

## I. INTRODUCTION

In complex-oxide molecular beam epitaxy (MBE) process, multiple source elements with significantly different oxygen affinities are used together[1]. However, the source oxidation of easily oxidizing elements leads to flux instability[2-3] and maintaining stable fluxes for all elements in an oxygen environment is a challenging task[4-5]. If all elements used are easily oxidized, such as in the case of Sr(Ca, Ba)TiO$_3$, a low background oxygen pressure, in $10^{-7}$ Torr range, suffices[6] and flux instability is not critical. However, this issue becomes prominent when an easily oxidized element is used together with a difficult to oxidize element, such as in cuprates[7-10] and PbTiO$_3$[3] that need a high pressure (~$10^{-5}$ Torr) ozone background to oxidize the difficult-to-oxidize elements – Cu in cuprates and Pb in PbTiO$_3$. Theis et. al. showed that while flux rate for Ti source remained fairly constant at a background ozone pressure of $2\times10^{-6}$ Torr, it dropped by 2.5% per hour when subjected to a ozone background pressure of $5\times10^{-5}$ Torr[11]. For elements like Ba, more than 50% flux drop has been observed in similar oxidation conditions[3]. In such harsh oxidation conditions, a real-time flux monitoring scheme such as atomic absorption spectroscopy (AA)[3,12-13] has to be employed to achieve a flux variation of less than 1% for the more easily oxidizing elements over several hours of growth. However, this increases the complexity of the growth process as the number of elements grows and is also cumbersome to implement. Minimizing oxygen partial pressure near the source surface even in a harsh oxygen environment is a fundamental solution for flux instability problem. Ideally, if the O$_2$ partial pressure near the source surface is negligible, the flux will be completely stable throughout the entire growth, thus eliminating the need for real-time monitoring.

In our previous reports, several parameters, such as port geometry, flux rate[14], and a new scheme involving a crucible aperture[15] were studied to enhance the flux stability against oxidation. It was found that extended port geometry enhanced the flux stability, both long term and short term, by lowering the oxygen conductivity through the source port. It was also observed that though the crucible aperture scheme considerably improves the flux stability, the exact position of the aperture was critical. Short term stability, in terms of oxygen pressure, was improved by placing the aperture near the crucible orifice, while long term stability was improved by placing it closer to the source surface[15]. Based on these results, we came up with a more powerful and effective scheme to enhance the flux stability, by using a collimator in an extended port geometry. The collimator minimizes the source area exposed to oxygen species like the crucible aperture and its long walls act as an effective oxygen getter like the extended port geometry. This self-gettering differential pump uses the source itself as a pumping medium to eliminate the source oxidation problem and stabilize the flux even when the flux is as low as ~ 0.01Å /sec, which is an order lower than the typical growth rate.

## II. EXPERIMENT

All the experiments were performed in a custom-designed SVTA MOS-V-2 MBE system with a base pressure of ~$10^{-10}$ Torr. High purity strontium (99.99 %) was loaded in a pyrolytic boron nitride (PBN) crucible and thermally evaporated from low-temperature effusion cells (SVTA-275/450/458-XX). The cell temperature was controlled by a Eurotherm 2408 temperature controller and the flux drift was less than 1% over several hours in an absence of oxygen ambient. The partial pressures of oxygen and Argon were controlled by a differentially-pumped mass flow controller (MFC) in combination with a precision leak valve. Again Sr was used as a test source, and the Sr flux was monitored using a quartz crystal microbalance (QCM).

Two source ports designed with different geometry were used; the standard port (STD) [Fig. 1(a)] which is typical in most MBE systems and the extended port (ETD) which is 21 cm longer than standard port [Fig. 1(c)]. One of the standard ports was modified for differential pumping using a port aperture (PA), mounted in front of the source, and connected to turbo pump on side of the source port [Fig. 1(b)]. The purpose of this differential pumping is to minimize the oxygen partial pressure near the source by mechanically pumping it through the turbo pump, as discussed later in detail. The port aperture with an inner diameter of 1.5 cm and a thickness of 3 mm is designed for easy insertion/removal using a few screws. A gate valve is inserted between the source port and the turbo pump to enable or disable differential pumping. In addition, an ion-gauge is mounted on the source port to read the gas pressure near the source. In the extended port geometry, a homemade collimator (inner diameter: 1.5 cm and length: 19 cm) is mounted in front of the source. We investigated flux stability against oxygen rich condition for all of these port geometries to find out the most effective configuration for minimizing source oxidation issues.

## III. RESULTS AND DISCUSSION

Short-term flux stability for the standard port and the extended port with various geometrical configurations, as discussed above, is shown in Fig. 2. These measurements were done in both oxygen and argon environment because the measured flux signals are affected not only by the source oxidation but also by the scattering effect[14]. From the flux scattering measurements conducted in the argon environment, it was found that the source scattering was dominated by the source-to-substrate distance (standard port; 21cm, extended port; 42cm) and not affected by other parameters such as flux rate, port aperture, collimator and differential pumping. This scattering process can be well described by the Beer–Lambert law[14]. Introducing oxygen gas into the chamber resulted in serious source oxidation in the standard port configuration and the flux decreased significantly at higher oxygen pressures [Fig. 2(a)]. After mounting a port aperture (PA) in front of the source, it was observed that the flux rate dropped from 0.15 to 0.06 Å/sec. Even at this low flux rate, an enhancement in the short term flux stability was observed. Using differential pumping through the turbo pump, connected to the back end of the source port, provided almost no additional improvement. In order to achieve similar flux rates to give a better comparison with the standard port, the source temperature was increased; it was observed that even though a significant improvement in flux stability was achieved at lower oxygen pressures, the problem still exists as the oxygen pressure is increased beyond $3 \times 10^{-6}$ Torr. The short-term flux stability for the extended port with/without collimator and differential pumping is shown in Fig. 2(b). At a similar flux rate, the extended port resulted in a comparable trend to that of a standard port with port aperture and the differential pumping. When the collimator is added to the extended port, interestingly, it was seen that the source was free from source oxidation issues even at an oxygen pressure as high as $5\times10^{-5}$ Torr. The short-term flux stability in this configuration was limited by the scattering of flux at higher pressures in a confined space along the collimator length, as seen by the measurements in Ar environment. At an $O_2$ pressure of $\sim 1\times10^{-5}$ Torr, the flux was seen to be fairly stable with flux rates, as low as 0.01 Å/sec, which is considerably lower than the typical growth rate in an MBE process.

To investigate long-term flux stabilities, we monitored both Sr flux and oxygen partial pressure in the source port for several hours, keeping oxygen pressure in the main chamber to be $1\times10^{-5}$ Torr. In a standard port without PA and DP, on introducing oxygen in the growth chamber, the oxygen pressure inside the source port increased suddenly; resulting in an abrupt, followed by a continuous, drop in the flux rate. A port aperture on the source port lowered the oxygen conductivity and the $O_2$ partial pressure in the source port increased slowly to its equilibrium value; a similar trend is observed for the drop in the flux rate[Fig. 3(a, c)]. It took ~30 minutes (~1.5 hours) for $O_2$ partial pressure and the flux rate to reach an equilibrium state with (without) differential pumping at the source port [Fig. 3(a, c)]. Beyond this, the oxygen pressure still increased without reaching a saturation point, causing the flux to decrease continuously. Considering that our goal is to make the flux drift less than 1% over several hours of growth, this long time delay to reach the equilibrium state and the continuous decrease of flux are not desirable. This result implies that even if the oxygen molecules in the source port were pumped out mechanically through the turbo pump, the oxygen partial pressure near the source was not low enough to prevent the source oxidation problem.

On the contrary, in the extended port with collimator, good long-term flux stabilities were observed, even without mechanical pumping; oxygen pressure near the source quickly saturated within minutes to an equilibrium value even at very low flux rate ~0.013 Å/sec, which is ~10 times lower than typical growth rate [Fig. 3(b, d)]. The flux had to be extremely low (below 0.009 Å/sec) to observe source oxidation, at a much reduced scale compared to a standard port with port aperture and differential pumping. This powerful enhancement in flux stability for a collimator mounted extended port can be explained in the following way: once the source is heated, the source atoms are deposited continuously on the wall of the collimator. Because of the long length and the large surface area of the collimator, most of the oxygen species coming into the source port stick to the collimator wall, due to chemical reaction with Sr (forming strontium oxide), without reaching the source itself. This gettering effect reduced the effective oxygen conductance through the collimator by a factor of a thousand compared to its geometric conductance for our chosen geometry (inner diameter: 1.5 cm and length: 19 cm); resulting in an effective oxygen partial pressure near the source well below the source oxidation regime and stabilizing the source flux over a long period of time, even at high oxygen pressures in the growth chamber. As shown in Fig. 4, the flux drift was negligible, less than 1 %, for three hours which is a typical operating time in an oxide-MBE process.

## IV. CONCLUSIONS

A self-gettering pump scheme has been proposed and evaluated in the context of the source oxidation problem in complex-oxide MBE process. It uses the Sr source itself as the oxygen pumping medium and reduced the effective oxygen conductance through the collimator by a factor of a thousand compared to its geometric conductance. As a result, the oxygen partial pressure near the source was easily maintained well below the source oxidation regime even when the growth chamber was in a harsh oxidation condition, and the source flux remained as stable as that of an ultra-high-vacuum environment. We were able to achieve a stable Sr flux even when the flux was much lower than typical growth rate in an oxygen pressure of $1 \times 10^{-5}$ Torr. We conceptually demonstrated this scheme using a collimator with an inner diameter: 1.5 cm and length: 19 cm, designed for uniform deposition over small samples (1cm x 1cm). A trade-off would be required between achieving this high level of source stability and deposition uniformity over large samples and the exact dimensions of the collimator would have to be optimized. However, for a collimator of this dimension, no sign of choking was observed on the collimator wall even after more than 100 hrs of operation, further demonstrating its effectiveness. This self-gettering pump scheme was tested with Sr, but it should work in a similar way for any other easily oxidizing elements and provide powerful solution for the source oxidation problem in oxide-MBE system.

# ACKNOWLEDGMENTS

This work is supported by IAMDN of Rutgers University, National Science Foundation (NSF DMR-0845464) and Office of Naval Research (ONR N000140910749). We thank Eliav Edrey for proofreading the manuscript.

# Figure Captions

**Fig. 1.** Cross sections of (a) standard port (STD), (b) STD with port aperture (PA) and differential pumping (DP), and (c) extended port (ETD) with collimator (COL).

**Fig. 2.** Short-term flux stability for (a) standard port with/without port aperture (PA) and differential pumping (DP) and (b) extended port with/without collimator (COL) and differential pumping (DP) at various flux rates.

**Fig. 3.** Time dependence of oxygen pressure near Sr source at (a) standard port and (b) extended port with various geometrical configurations when the oxygen partial pressure in main chamber is kept $1 \times 10^{-5}$ Torr. Long term flux stability of (c) standard port and (d) extended port.

**Fig. 4.** Normalized long-term flux stability for the extended port with collimator and the standard port with/without port aperture, differential pumping. In extended port with collimator, the flux variation over three hours is less 1% even if the flux rate is much slower than typical growth rate while that of standard port is more than 2% even with higher growth rate. The standard port with PA and DP shows the worst performance; over three hours, flux drops by more than 10%, implying any improper differential pumping is worse than having none.

# Figures

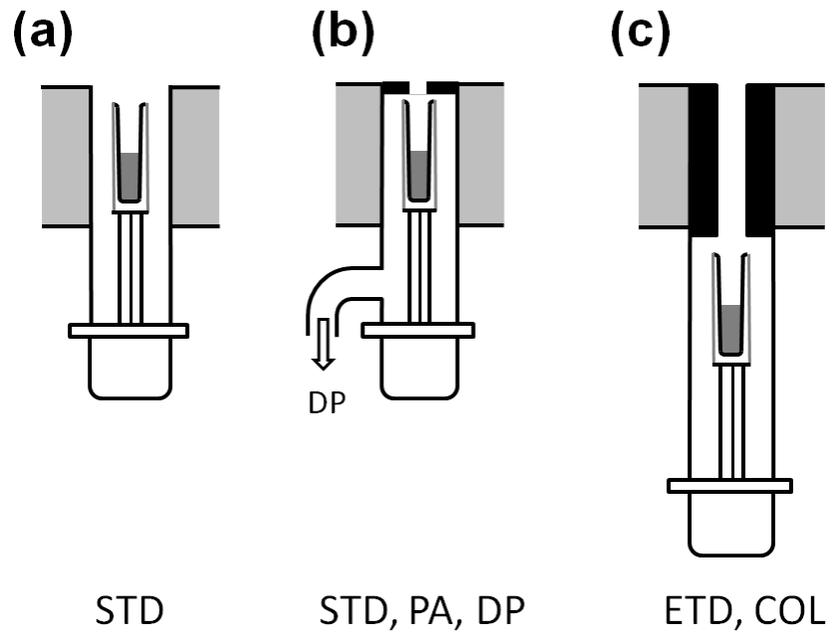

Fig. 1.

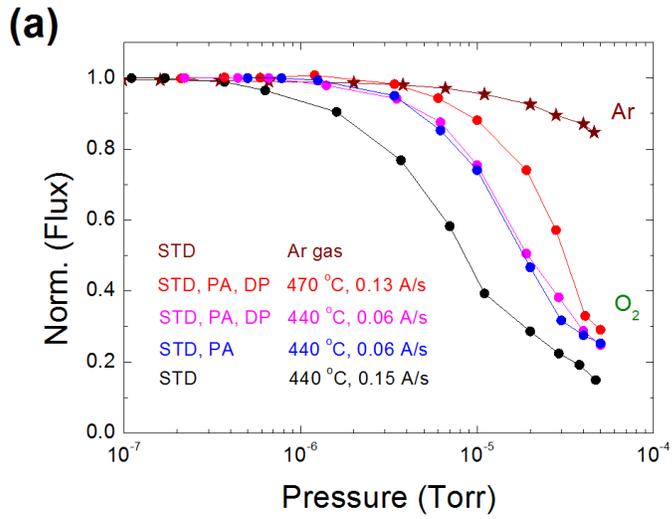
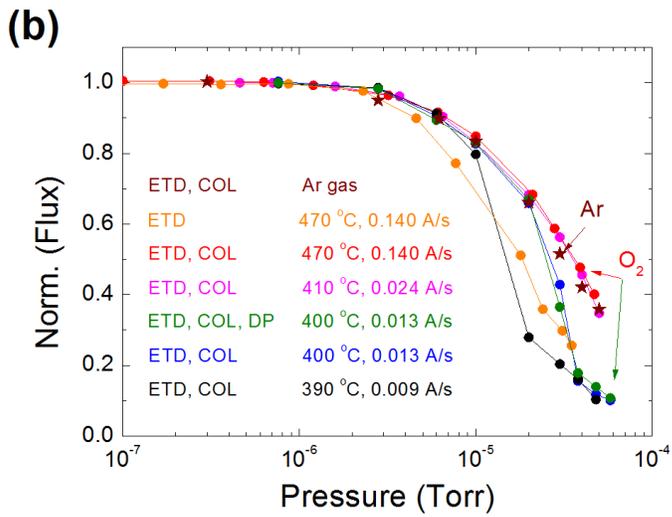

**Fig. 2.**

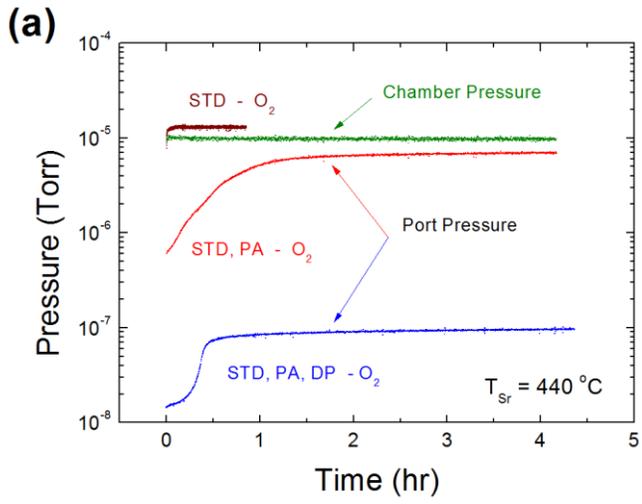
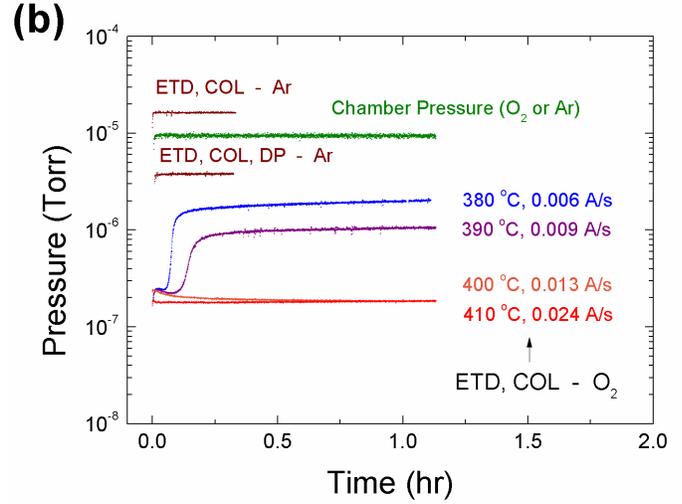
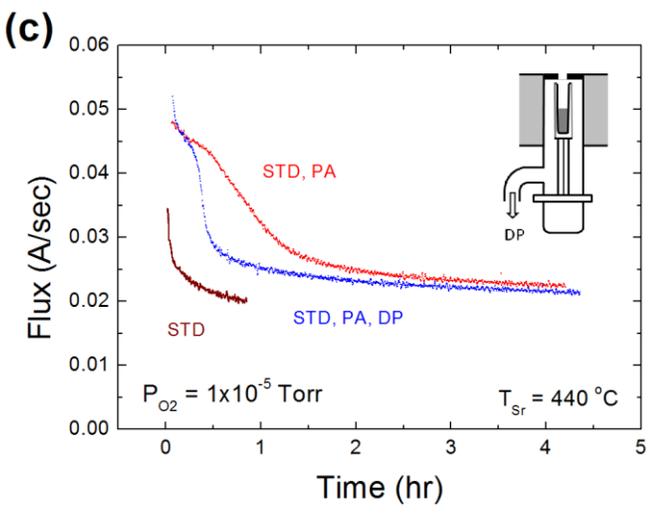
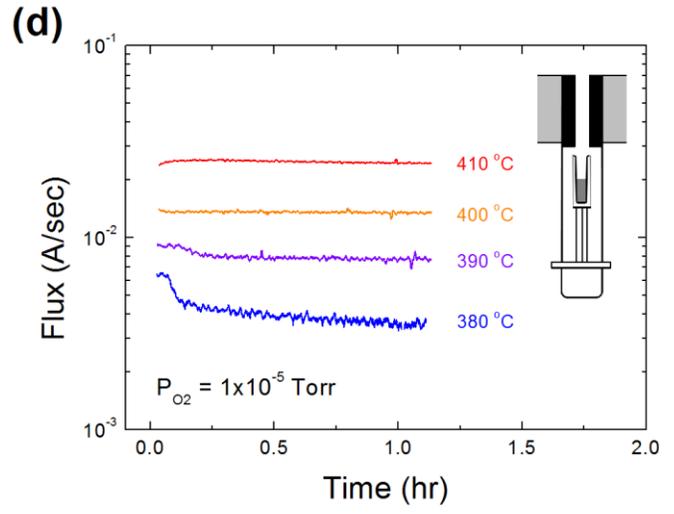

**Fig. 3.**

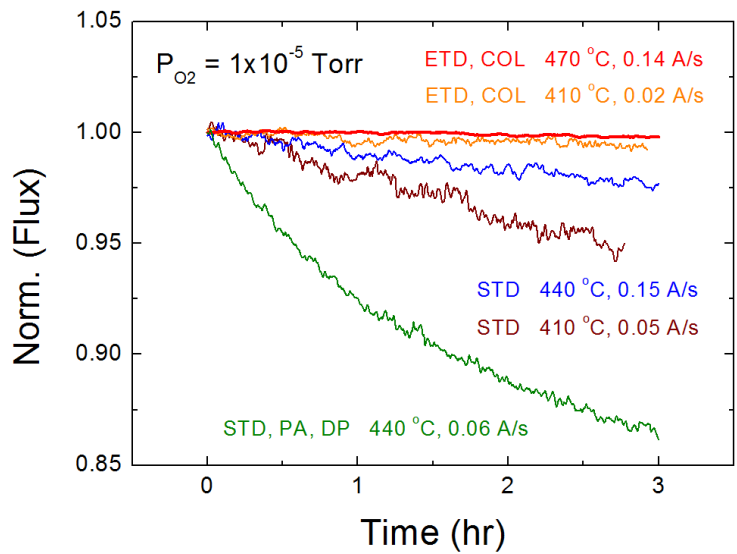

**Fig. 4.**